\documentclass[5p]{elsarticle}
\usepackage{array}
\usepackage{color}
\usepackage{multirow}
\usepackage{dcolumn}
\usepackage{bm}
\usepackage{epsf}
\usepackage{graphicx}
\usepackage{amsmath} 
\usepackage{amssymb}
\usepackage{dsfont} 
\usepackage{tikz}
\usepackage{hyperref}

\begin{document}

\begin{frontmatter}

\title{SN1987A and neutrino non-radiative decay}

\author[label1]{Pilar Iv\'a$\tilde{\rm n}$ez-Ballesteros}
\author[label2]{Maria Cristina Volpe}

\affiliation[label1]{organization={Universite Paris Cite, Astroparticule et Cosmologie},
            addressline={10, rue A. Domon et L. Duquet}, 
            city={Paris},
            postcode={ F-75013 }, 
            country={France}}

\affiliation[label2]{organization={CNRS, Universite Paris Cite, Astroparticule et Cosmologie},
            addressline={10, rue A. Domon et L. Duquet}, 
            city={Paris},
            postcode={ F-75013 }, 
            country={France}}

\begin{abstract}
We investigate neutrino non-radiative two-body decay in vacuum, in relation to SN1987A.
In a full $3\nu$ decay framework, we perform a detailed likelihood analysis of the 24 neutrino events from SN1987A observed by Kamiokande-II, IMB, and Baksan.
We consider both normal and inverted neutrino mass orderings, and the possibility of strongly hierarchical and quasi-degenerate
neutrino mass patterns. The results of the likelihood analysis show that the sensitivity is too low to derive bounds in the case of normal mass ordering. On the contrary, in the case of inverted mass ordering we obtain the bound $\tau/m \ge 2.4 \times 10^{5}$ s/eV ($1.2 \times 10^{5}$) s/eV at 68 $\%$ (90 $\%$) CL on the lifetime-to-mass ratio of the mass eigenstates $\nu_2$ and $\nu_1$.
\end{abstract}


\end{frontmatter}

\section{Introduction}
\label{Introduction}

SN1987A was a unique event, observed in all wavelengths. It was the first time that neutrinos were observed from the core collapse
of a massive star. On the 23$^{rd}$ of February 1987 the explosion of blue supergiant Sanduleak-69$^{\circ}$202, in the Large Magellanic Cloud, was observed.
The Kamiokande-II \cite{Kamiokande-II:1987idp}, the IMB \cite{Bionta:1987qt}, and the Baksan \cite{Alekseev:1988gp} detectors recorded 11, 8, and 5 neutrino events, respectively.
The Mont Blanc liquid scintillator detector \cite{Aglietta:1987it} detected 5 events, 5 hours earlier, that remain debated. Since this pioneering observation,
SN1987A has been a marvelous laboratory for astrophysics and particle physics. 

While neutrino vacuum oscillations are well established \cite{Super-Kamiokande:1998kpq,SNO:2001kpb,KamLAND:2002uet},
how neutrinos change flavor in dense environments (e.g. core-collapse supernovae and binary neutron star mergers) requires further investigations (see \cite{Volpe:2023met, Duan:2010bg, Mirizzi:2015eza, Tamborra:2020cul} for a comprehensive review). The neutrino mixing angles, relating the flavor
to the mass basis, are precisely measured (although the $\theta_{23}$ octant remains unknown) \cite{ParticleDataGroup:2022pth}. Moreover, global analysis of oscillation data
shows hints at 2.5$\sigma$ for normal mass ordering and for $\sin \delta < 0$ at 90\% confidence level (CL) for the Dirac CP-violating phase \cite{Capozzi:2021fjo, Gonzalez-Garcia:2014bfa, Esteban:2018azc}.  
As for the absolute neutrino mass, the KATRIN experiment put the upper limit $m < 0.8$ eV at 90\% CL
from tritium $\beta$ decay \cite{KATRIN:2021uub}. Cosmological observations have been able to constrain the sum of the neutrino masses; using Planck2018 temperature and polarization data: $\sum m_\nu < 0.26$~eV (95\% CL)~\cite{Planck:2018vyg}.

\begin{figure}[b!]
\begin{center}
\includegraphics[scale=0.45]{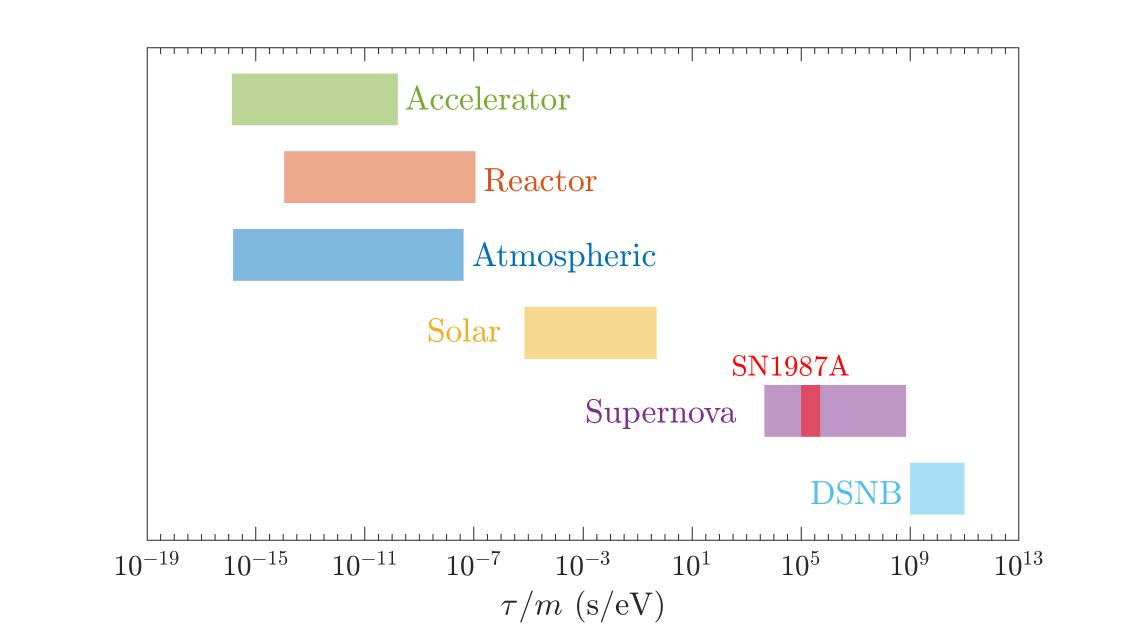}
\caption{Typical sensitivities to lifetime-to-mass ratios for neutrino non-radiative decay for different neutrino sources. The bands are obtained by considering typical distances and neutrino energies for the corresponding experiments and considering either almost full decay ($1 \%$ of the initial flux) or practically no decay ($99 \%$ of the initial flux). Also the band corresponding to SN1987A is shown, corresponding to $E_{\nu} \in [10, 50]$ MeV.}
\label{fig:taumtyp}
\end{center}
\end{figure}

Interesting neutrino properties remain unknown, e.g. the neutrino magnetic moment (see \cite{Giunti:2014ixa} for a review),
or neutrino decay, which was first evoked by Bahcall, Cabibbo, and Yahil \cite{Bahcall:1972my}.
In the presence of decay, the flux of neutrinos of mass $m$ and energy $E$ at a distance $L$ decreases as $\exp{(- L/E \times m/\tau)}$. From this suppression, one can get limits for the lifetime-to-mass ratio, $\tau/m$. Figure~\ref{fig:taumtyp} shows the sensitivity to $\tau/m$ for different neutrino sources considering the different characteristic baselines and neutrino energies.

Numerous studies yielded lower bounds on non-radiative two-body decay, by
considering that the final states involve either only invisible particles or
an invisible particle and an active neutrino. For example, assuming normal
mass ordering, Ref.~\cite{Gonzalez-Garcia:2008mgl} obtained the limit $\tau
_{3}/m_{3} > 10^{-11}$~s/eV at 99\% CL from the oscillation plus decay analysis of
combined atmospheric and accelerator data. Ref.~\cite{Beacom:2002cb}
considered the impact of non-radiative decay on solar neutrinos either to
sterile species or to active neutrinos and a light (pseudo)scalar boson, and
derived the lower bound of $\tau _{2} / m_{2} > 3 \times 10^{-4}$~s/eV. From
SNO and other solar neutrino data, the constraint $\tau _{2} / m_{2} > 1.04
\times 10^{-3}$~s/eV was obtained at 99\% CL \cite{SNO:2018pvg}. Using
Planck2018 data, Ref.~\cite{Barenboim:2020vrr} derived the limit $\tau > 4
\times 10^{5}\text{--}10^{6}~(m/0.05~{\text{eV}})^{5}~\text{s}$ for decay
into massless neutrinos, whereas the bound becomes weaker for decay into
massive neutrinos \cite{Chen:2022idm}. Moreover, Ref.~\cite{Beacom:2002vi}
anticipated unique signatures for neutrino non-radiative decay with
ultra-high energy neutrinos from distant sources. Indeed their observation by
the IceCube Collaboration \cite{IceCube:2013low} yielded limits of $\tau / m
> 10^{3}$~s/eV at 95\% CL (for the case of complete decay) \cite{Pagliaroli:2015rca}.
Ref.~\cite{Denton:2018aml} found a hint for invisible neutrino decay with
$\tau /m = 10^{2}$~s/eV. This possibility was further studied in
Ref.~\cite{Abdullahi:2020rge} where visible decay was considered as well.
 
Future observations of the diffuse supernova neutrino background, from past core-collapse supernovae, have a unique sensitivity in the range $\tau/m \in [10^{9}, 10^{11}]$ s/eV \cite{Fogli:2004gy,Tabrizi:2020vmo,DeGouvea:2020ang}. 
Ref. \cite{Ivanez-Ballesteros:2022szu} performed a full $3 \nu$ flavor analysis of neutrino non-radiative decay, taking into account both astrophysical uncertainties from the evolving core-collapse supernova rate and the fraction of failed supernovae. The results showed, for both mass orderings, important degeneracies between the expected rates with no-decay and standard physics inputs, and the ones including neutrino decay.  

Clearly, future astrophysical observations will bring crucial information to push current limits to larger $\tau/m$. 
If a supernova blows off in our galaxy, neutrino decay can impact the neutronization burst, for which DUNE and Hyper-Kamiokande will have a sensitivity
for $\tau/m < 10^{6}$-$10^{7}$ s/eV respectively \cite{deGouvea:2019goq}. Concerning SN1987A, tight limits were obtained for neutrino radiative decay \cite{Raffelt:1996wa}. In Ref.~\cite{Fiorillo:2022cdq} the authors obtained constraints on Majoron-neutrino couplings from the study of SN1987A data: $g \lesssim 10^{-9}$~MeV$/m_\phi$ for a Majoron mass $m_{\phi} \in [100~{\rm eV} -100~{\rm MeV}]$.
On the other hand, Refs. \cite{Kachelriess:2000qc,Farzan:2002wx} considered Majoron models and the possibility that neutrino non-radiative decay in matter influences supernova cooling. 
In contrast, Ref. \cite{Frieman:1987as} studied an oscillation plus decay scenario in vacuum, relating their findings to the solar neutrino problem. 

In this work, we present a detailed investigation of neutrino non-radiative two-body decay, using the 24 events from SN1987A.
First, with a two-degrees-of-freedom (2D) likelihood analysis, we obtain 
the best fits and allowed regions for the electron antineutrino average energy and luminosity in the absence of decay.
Then, based on a full $3 \nu$ flavor framework, we calculate the supernova neutrino fluxes in the presence of decay, for normal and inverted mass orderings.  
We perform a seven-degrees-of-freedom (7D) likelihood analysis of the Kamiokande-II, IMB, and Baksan data including neutrino non-radiative decay. We show their sensitivity 
to neutrino decay in inverted and in normal neutrino mass ordering for the strongly hierarchical and quasi-degenerate mass patterns.  
Finally, we present our bounds on the lifetime-to-mass ratio. 

The present manuscript is structured as follows. First, in Section~2
we present the theoretical approach used to derive the supernova neutrino
fluxes and include neutrino decay. In Section~3, we introduce the
calculation of the signal in Kamiokande-II, IMB, and Baksan. Section~4 presents the likelihood and the test statistic used in this work.
In Section~5 we first show the numerical results on the $\bar{\nu}_{e}$
average energy and luminosity in the absence of decay. Then we give our
bounds on the lifetime-to-mass ratio for non-radiative neutrino decay for the different cases considered and discuss the results obtained. In
Appendix~A, we give the explicit equations used to determine the
neutrino fluxes. Appendix~B provides SN1987A data, for consistency,
from the Kamiokande-II, IMB, and Baksan experiments.

\section{Supernova neutrinos in presence of neutrino non-radiative decay}
\subsection{Neutrino fluxes before decay}
Neutrino fluences (time-integrated fluxes) produced in a supernova core are well represented by power-law distributions \cite{Keil:2002in}
\begin{equation}\label{eq:PW}
 \phi^0_{\nu}(E_{\nu}) = {(\alpha + 1)^{\alpha + 1} \over{\langle E_{\nu} \rangle \Gamma(\alpha + 1)}}  \Big({ E_{\nu} \over {\langle E_{\nu} \rangle}} \Big)^{\alpha}{\rm  e}^{ -{ (1+ \alpha) E_{\nu} \over{\langle E_{\nu}} \rangle }} \ , 
\end{equation}
and the corresponding neutrino yields are 
\begin{equation}\label{eq:yield}
F^0_{\nu} =  {L_{\nu} \over{\langle E_{\nu} \rangle}}  \phi^0_{\nu}(E_{\nu})  \ ,
\end{equation}
with $\langle E_{\nu} \rangle$ the average neutrino energy, $\alpha$ the pinching parameter and $L_{\nu}$ the neutrino luminosity that corresponds to $99 \%$ of the gravitational binding energy emitted by the supernova. 

When neutrinos travel through the supernova layers, they interact with the background particles and undergo flavor conversion due to the established Mikheev-Smirnov-Wolfenstein effect
\cite{Wolfenstein:1977ue,Mikheev:1986wj}. This phenomenon produces spectral swapping for neutrinos according to \cite{Dighe:1999bi}
\begin{equation}\label{eq:MSWnu}
F_{{\nu}_l}  = F^0_{{\nu}_x} ~~F_{{\nu}_i} = F^0_{\nu_x} ~~F_{{\nu}_h} = F^0_{\nu_e}  \ , 
\end{equation}
and for antineutrinos
\begin{equation}\label{eq:MSWnubar}
F_{\bar{\nu}_l}  = F^0_{\bar{\nu}_e}~~ F_{\bar{\nu}_i} = F^0_{\bar{\nu}_x} ~~F_{\bar{\nu}_h} = F^0_{\bar{\nu}_x}  \ ,
\end{equation}
where these relations hold if the neutrino mass ordering is normal (NO) (i.e. $\Delta m^2_{31} > 0$); or inverted (IO) (i.e. $\Delta m^2_{31} < 0$). 
Here $\nu_h$ refers to the heaviest, $\nu_i$ to the intermediate, and $\nu_l$ to the lightest mass eigenstates. 

\subsection{Neutrino fluxes when neutrinos decay}
Once neutrinos reach the supernova surface, they can decay in their path to the Earth. Note that we do not consider the possibility that neutrinos decay in matter, previously investigated in Refs.\cite{Kachelriess:2000qc,Farzan:2002wx}. Moreover, we assume (as commonly done) that the mass and the decaying eigenstates coincide.
The effects of their mismatch were discussed in Refs. \cite{Berryman:2014qha,Chattopadhyay:2021eba}. 

We consider that a heavier neutrino, $\nu_h$, can decay to a lighter one, $\nu_l$, and a massless 
(pseudo)scalar boson, $\phi$, such as the Majoron \cite{Chikashige:1980ui,Gelmini:1980re}, which does not carry definite lepton number (see e.g. \cite{Kim:1990km}),
\begin{equation}
\nu_h \rightarrow \nu_l + \phi   ~~~ {\rm or} ~~~  \nu_h \rightarrow \bar{\nu}_l + \phi  \ .
\end{equation}
The decay is due e.g. to tree-level scalar or pseudoscalar couplings of the form
\begin{equation}\label{eq:lagrangian}
{\cal L} = g_{ij} \bar{\nu}_i \nu_j \phi + h_{ij} \bar{\nu}_i \gamma_5 \nu_j \phi + H.c. \ , 
\end{equation}
where $\nu$ here are the neutrino fields; $\phi$, the new scalar field; and $g_{ij}$ and $h_{ij}$, the couplings of the neutrino fields to the new scalar field. We will not consider any specific model and keep the discussion general.

In the presence of neutrino non-radiative two-body decay the supernova neutrino fluxes on Earth are given by the transfer equations \cite{Ando:2004qe}
\begin{align}\label{eq:transf}
{d \over {dr}} F_{\bar{\nu}_k} (r, E_{\nu}) & = - \Gamma_{k} (E_{\nu})F_{\bar{\nu}_{k}} (r, E_{\nu})  \nonumber \\
+ \sum_{m_j > m_k} \int_{E_\nu}^{\infty} dE_{\nu}'  & [ \psi_{\rm h.c.}(E_{\nu}', E_{\nu}) \Gamma_{\bar{\nu}_j \rightarrow \bar{\nu}_k}(E'_{\nu}) F_{\bar{\nu}_j} (r,E_{\nu}')    \nonumber \\
 +  & \psi_{\rm h.f.}(E_{\nu}', E_{\nu}) \Gamma_{\nu_j \rightarrow \bar{\nu}_k}(E_{\nu}') F_{{\nu}_j} (r,E_{\nu}') ] \ ,
\end{align}
where
\begin{align}\label{eq:rate}
\Gamma_{k} = \sum_{m_j < m_k} \Gamma_{\bar{\nu}_k \rightarrow \bar{\nu}_j} + \Gamma_{\bar{\nu}_k \rightarrow \nu_j}
\end{align}
is the decay rate in the laboratory frame
and $r$ is the distance from the surface of the supernova. In Eq.~\ref{eq:rate}, $ \Gamma_{\bar{\nu}_k \rightarrow \bar{\nu}_j} $ and $\Gamma_{\nu_k \rightarrow \bar{\nu}_j}$ are the partial decay rates.
The functions $\psi_{\rm h.c.}$ and $\psi_{\rm h.f.}$ in Eqs.~\ref{eq:transf} indicate the energy decay spectrum for the helicity conserving (h.c.) and the helicity flipping (h.f.) decays, respectively.
The analytical solutions to the differential equations \eqref{eq:transf} are given in Appendix A. 

Equations \eqref{eq:transf} include both the depletion due to the decay of the heaviest neutrino states into the lighter ones (first term), and the flux increase 
of the intermediate and lightest neutrino states due to the decay of the heavier states. This increase includes the contribution from the helicity conserving (h.c.) and the helicity flipping (h.f.) decays (second and third terms on the {\it r.h.s.}).

The lifetime in the laboratory frame is
\begin{equation}\label{eq:lifetimelab}
\tau^{\rm lab}_{k} = \Gamma_k^{-1} (E) = { E_{\nu} \over m_k} \tau_{k}  \ , 
\end{equation}
with $\tau_{k}$ the lifetime in the rest frame. Note that
bounds on neutrino non-radiative decay are usually given for the lifetime-to-mass ratio since the absolute neutrino mass is not known yet.

\begin{figure*}
\begin{center}
\includegraphics[scale=0.5]{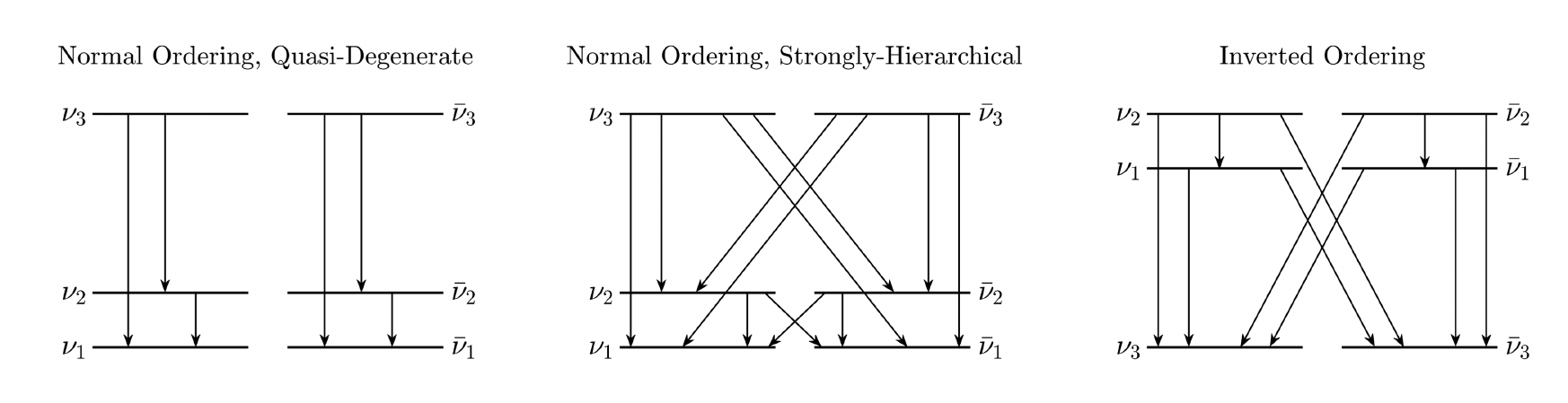}
\caption{Decay patterns for $3 \nu$ flavors. Left:  Normal mass ordering in the quasi-degenerate case (QD). The branching ratios are equal to 1/2 for $\nu_3$ or $\bar{\nu}_3$. 
Middle: Normal mass ordering in the strongly-hierarchical (SH) case. The branching ratios are equal to 1/4 for $\nu_3$ ($\bar{\nu}_3$) and 1/2 for $\nu_2$ ($\bar{\nu}_2$).  Right:  Inverted mass ordering case. The branching ratios are equal to 1/2 for $\nu_2 \to \nu_1$ ($\bar{\nu}_2 \to \bar{\nu}_1$), 1/4 for $\nu_2 \to \overset{(-)}{\nu_3}$ ($\bar{\nu}_2 \to \overset{(-)}{\nu_3}$) and 1/2 for $\nu_1$ ($\bar{\nu}_1$).
In all cases the lifetime-to-mass ratio of the decaying eigenstates is taken equal, i.e. $\tau_2/m_2 = \tau_3/m_3 $ (NO) or $\tau_2/m_2 = \tau_1/m_1 $ (IO).}
\label{fig:decaypatterns}
\end{center}
\end{figure*}

In order to solve Eqs.\eqref{eq:transf} one needs to define the neutrino mass pattern.
We consider here two extreme cases:
\begin{itemize}
\item  $m_h - m_l \gg m_l \simeq 0$ ---  strongly hierarchical (SH);
\item $m_h \simeq m_l \gg m_h - m_l$ --- quasi-degenerate (QD).
\end{itemize} 
 Figure \ref{fig:decaypatterns} presents the cases of NO and SH or QD as well as of IO. The latter comprises a 
QD pattern for $m_1$, $m_2$ which are strongly hierarchical with respect to $m_3$.
We made a {\it democratic ansatz} on the branching ratios for the decaying eigenstates and we assumed that their lifetime-to-mass ratios are equal so that
there is only one free parameter to determine.

The energy-dependent functions, that account for the spectral distortions in Eqs.\eqref{eq:transf}, depend strongly on the mass
pattern. More precisely, in the QD case in which only helicity conserving decay is allowed, the daughter neutrino inherits nearly the full energy of the parent neutrino
\begin{equation}\label{eq:QD}
\psi (E_{\nu_h}, E_{\nu_l}) = \delta(E_{\nu_h} -E_{\nu_l}) \ .
\end{equation} 
On another hand, in the SH case one has that
\begin{equation}\label{SH}
\psi_{\rm h.c.} (E_{\nu_h}, E_{\nu_l}) = { 2 E_{\nu_l} \over{E_{\nu_h}^2}} ~~~~\psi_{\rm h.f.} (E_{\nu_h}, E_{\nu_l}) =  {2 \over {E_{\nu_h}}} \Big(1 - { E_{\nu_l} \over {E_{\nu_h}}}   \Big) \ ,
\end{equation}
so that daughter neutrinos from helicity-conserving decay have higher energy than the one of those coming from helicity-flipping decay.

The $\bar{\nu}_e$ flux at the detector is given by
\begin{equation}
F_{\bar{\nu}_e} (E_{\nu}, L) = \sum_k \vert U_{ek}  \vert^2 F_{\bar{\nu}_k} (E_{\nu}, L) 
\end{equation}
where $U_{ek} $ are matrix elements of the Pontecorvo-Maki-Nakagawa-Sakata matrix for which we use
$\theta_{23} = 45^{\circ}, \theta_{12} = 34^{\circ},\theta_{13} = 8.5^{\circ}$ \cite{ParticleDataGroup:2022pth}.
The antineutrino yields include flavor conversion due to the MSW mechanism in the supernova envelope \eqref{eq:MSWnubar} and the
appearance and disappearance terms of the transfer equations \eqref{eq:transf}  for which there are analytical solutions (Appendix A). 
For the flux suppression due to the inverse squared distance $L^{-2}$, we take $L = 50$ kpc.

\section{The neutrino signal from SN1987A}
We describe here the most important ingredients to determine the neutrino signal in Kamiokande-II, IMB, and Baksan. 

The main detection channel in the three detectors was inverse beta-decay (IBD)
\begin{equation}\label{IBD}
\bar{\nu}_e + p \rightarrow n + e^+  \ .
\end{equation}
For the event calculations, we employ the 
cross section from \cite{Strumia:2003zx} (see also Ref.\cite{Ricciardi:2022pru}). Note that the positron energy is $E_{e} = E_{\bar{\nu}_e} - \Delta_{np} $, where $\Delta_{np} = m_n - m_p = 1.293$~MeV is the difference between the neutron and proton masses, and the energy threshold of the IBD process is 1.806 MeV.

In the description of the signal, we follow Ref. \cite{Vissani:2014doa}. 
We describe the spectrum of the positron emitted in the IBD process using the approximate expression
\begin{equation}\label{eq:pos}
  \frac{d S_e}{d E_e} \approx N_p F_{\bar\nu_e}(E_\nu) \sigma_\mathrm{IBD}(E_\nu) J(E_{\nu}) \ ,
  \end{equation} 
that describes well the IBD signal in the energy range relevant for supernova detection.  
In Equation \eqref{eq:pos} one has that $N_p$ is the number of target protons at each detector (see Table~\ref{tab:properties}), $\sigma_{\rm IBD}$ is the IBD cross-section, and
\begin{equation}    
E_\nu = \frac{E_e + \Delta_{np}}{1 - E_e / m_p} \ . 
\end{equation} 
The term $J(E_\nu)$ is the Jacobian factor which has the following form 
\begin{equation}
    J(E_\nu) = \frac{\left( 1+ E_\nu / m_p\right)^2}{1 + \Delta_{np} / m_p}.
\end{equation}

The observed positron spectrum can be related to the true positron spectrum through 
\begin{equation}
    \frac{d S}{d E_i} = \int^\infty_0 \eta(E_e) G\big(E_e - E_i, \sigma(E_e) \big) \frac{d S_e}{d E_e} d E_e ,
    \label{eq:signal_obs}
\end{equation}
where $\eta(E_e)$ is the \textit{intrinsic efficiency function} of the detectors, $G$ the \textit{smearing function} assumed to be Gaussian.
The quantity $E_i$ is the reconstructed positron energy for a given true energy $E_e$. 

The smearing function depends on the uncertainty function 
\begin{equation}
    \sigma (E_e) = \sigma_\mathrm{stat} \times \left( \frac{E_e}{10~\mathrm{MeV}} \right)^{1/2} + \sigma_\mathrm{syst} \times \left( \frac{E_e}{10~\mathrm{MeV}} \right).
\end{equation}
Table~\ref{tab:properties} gives the two coefficients, $\sigma_\mathrm{stat}$ and $\sigma_\mathrm{syst}$, of its statistical and systematic components. 
These were determined by Ref.~\cite{Vissani:2014doa} using the observed energies and related error
in each experiment.

\begin{table}[t]
\begin{center}
\begin{tabular}{l|cccc|}
\cline{2-5}
 & \multicolumn{1}{c}{$N_p $} & \multicolumn{1}{c}{\begin{tabular}[c]{c}$\sigma_{\rm stat}$\\ {[}MeV{]}\end{tabular}} & \multicolumn{1}{c}{\begin{tabular}[c]{c}$\sigma_{\rm syst}$\\ {[}MeV{]}\end{tabular}} & \multicolumn{1}{c|}{$B_{\rm tot}$} \\ \hline
\multicolumn{1}{|l|}{Kamiokande-II}   & $1.4 \times 10^{32}$      & 1.27  & 1.0   & 0.55     \\ \hline
\multicolumn{1}{|l|}{IMB}    & $4.6 \times 10^{32}$      & 3.0   & 0.4   & 0.01      \\ \hline
\multicolumn{1}{|l|}{Baksan} & $0.2 \times 10^{32}$      & 0.0    & 2.0  & 1.0        \\ \hline
\end{tabular}
\end{center}
\caption{Properties of the experiments (first column) that detected the neutrinos from SN1987A. The table shows the number of target protons (second column) and the two coefficients defining the uncertainty function $\sigma(E_{e})$ (third and fourth columns). The last column gives the total background $B_{\rm tot}$ (see text). (From \cite{Vissani:2014doa}.)}\label{tab:properties}
\end{table}

Finally, the total number of expected signal events above threshold, $E_\mathrm{min}$, can be obtained using the true positron spectrum through the following expression:
\begin{equation}\label{eq:tot}
    S_\mathrm{tot} = \int_0^\infty \eta(E_e)  g(E_e, E_\mathrm{min}) \frac{d S_e}{d E_e} d E_e,
\end{equation}
with \footnote{The function Erf is defined as $\int^y_0 G(x, \sigma) dx = \frac{1}{2}\mathrm{Erf}\left[ \frac{y}{\sqrt{2}\sigma}\right]$.}
\begin{equation}
    g(E_e, E_\mathrm{min}) = \frac{1}{2} \Big[1 + \mathrm{Erf} \left( \frac{E_e - E_\mathrm{min}}{\sqrt{2} \sigma (E_e)}\right)\Big] \ , 
\end{equation} 
and $ \eta(E_e)  g(E_e, E_\mathrm{min}) $ the total efficiency.

The last term to be defined in Eqs.~(\ref{eq:signal_obs}) and \eqref{eq:tot} is $\eta(E_e)$ which depends on the detector. 
The parameterizations deduced by Ref.~\cite{Vissani:2014doa} 
from published data strongly deviates from 1 except for Baksan. 
For Kamiokande-II, we use the intrinsic efficiency 
\begin{equation}\label{eq:intrinsic_KII}
    \begin{split}
        \eta (E_e) = 0.93 \Bigg[\Bigg. 1 - \left( \frac{0.2~\mathrm{MeV}}{E_e}\right)& - \left( \frac{2.5~\mathrm{MeV}}{E_e}\right)^2 \Bigg.\Bigg]
        \\ & \mathrm{for } ~ E_e > E_\mathrm{KII} \ .    
    \end{split}     
\end{equation}
For IMB we employ
\begin{equation}\label{eq:intrinsic_IMB}
    \begin{split}
        \eta (E_e) = \sum_{n=1}^5 c_n(\mathrm{IMB})&\times \left( \frac{E_e}{E_\mathrm{IMB}} -1\right)^n 
        \\ & \mathrm{for } ~ E_\mathrm{IMB} < E_e < 70~\mathrm{MeV},
    \end{split}
\end{equation}
for which $c_1(\mathrm{IMB}) = 0.369$, $c_4(\mathrm{IMB}) = -6 \times 10^{-4}$, $c_5(\mathrm{IMB}) = 10^{-4}$, and $c_{j \neq 1, 4, 5} = 0$. Including uncertainties, the IMB energy threshold is $E_\mathrm{IMB} = 15 \pm 2$~MeV.
For Baksan, we take 
\begin{equation}\
\eta (E_e) = 1~~\forall E_e \ .
\end{equation}

In our calculations, we use the energy thresholds $E_\mathrm{IMB} = 15$~MeV and $E_\mathrm{B} = 10$~MeV for IMB and Baksan experiments, respectively. 
Finally, we take the value $E_\mathrm{KII} = 7.5$~MeV, commonly employed to discriminate the signal from the background in the events reported by Kamiokande-II.

\section{Likelihood analysis}
Since we are considering a small data set, for the statistical analysis, we use an unbinned likelihood \cite{Vissani:2014doa} 
\begin{equation} \label{eq:lh}
\mathcal{L}(x)=e^{-(S_{\mathrm{tot}}(x) + B_{\mathrm{tot}})} \times \prod_{i=1}^{n_ \mathrm{obs}} \mathrm{d}E\left[ \frac{\mathrm{d}S}{\mathrm{d}E_i}(x) + \frac{\mathrm{d}B}{\mathrm{d}E_i} \right],
\end{equation}
where $x$ represents the set of parameters from our model. 
The total number of events is given by Eq. \eqref{eq:tot}, while the total background $B_{\mathrm{tot}}$
is obtained from the last column of Table~\ref{tab:properties}.
The product in Eq.\eqref{eq:lh} contains $n_\mathrm{obs}$ factors, with $n_\mathrm{obs}$ the total number of observed events. 
The first term 
in brackets gives the expected number of signal events around the observed energy $E_i$, while the second term is the background rate. The values of the background rate, ${\rm d}B/{\rm d}E_i$, are obtained by multiplying the values in Table~\ref{tab:SN1987Aevents} by a signal duration that we take to be 30 s.

To test values of the lifetime-to-mass ratio we consider the profile likelihood ratio  
\begin{equation} \label{eq:proflh}
    \lambda (\tau/m) = \frac{\mathcal{L}_p}{\mathcal{L}_{max}} = \frac{\mathcal{L} (\tau/m, \hat{\hat{\mathbf{L}}}, \hat{\hat{\mathbf{\langle E \rangle}}})}{\mathcal{L} (\hat\tau/m, \hat{\mathbf{L}}, \hat{\mathbf{\langle E \rangle}})}.
\end{equation}
where $\mathcal{L}_p$ is the profile likelihood that maximizes the likelihood for a given $\tau/m$. The quantity $\mathcal{L}_{max}$ is the unconditional maximized likelihood function.

Using Eq.~(\ref{eq:proflh}), we can define the test statistic as
\begin{equation}\label{eq:tauom}
    \chi^2 = -2 \log \lambda (\tau/m) \ , 
\end{equation}
i.e. it follows a $\chi^2$ distribution (for sufficiently large samples) \cite{Cowan:2010js}. Using Eqs.\eqref{eq:proflh}-\eqref{eq:tauom} we determine information on the lifetime-to-mass ratio. The results of the likelihood analysis are obtained using the simplex method. 

\section{Numerical results}
We now introduce the outcome of our investigation of SN1987A events. 
For consistency, Table \ref{tab:SN1987Aevents} summarizes the relevant information on the observed neutrino events used in our analysis, namely the positron energies and the associated background rates in Kamiokande-II, IMB and Baksan detectors.

\subsection{Results in absence of decay}
\begin{figure}
\begin{center}
\includegraphics[scale=0.4]{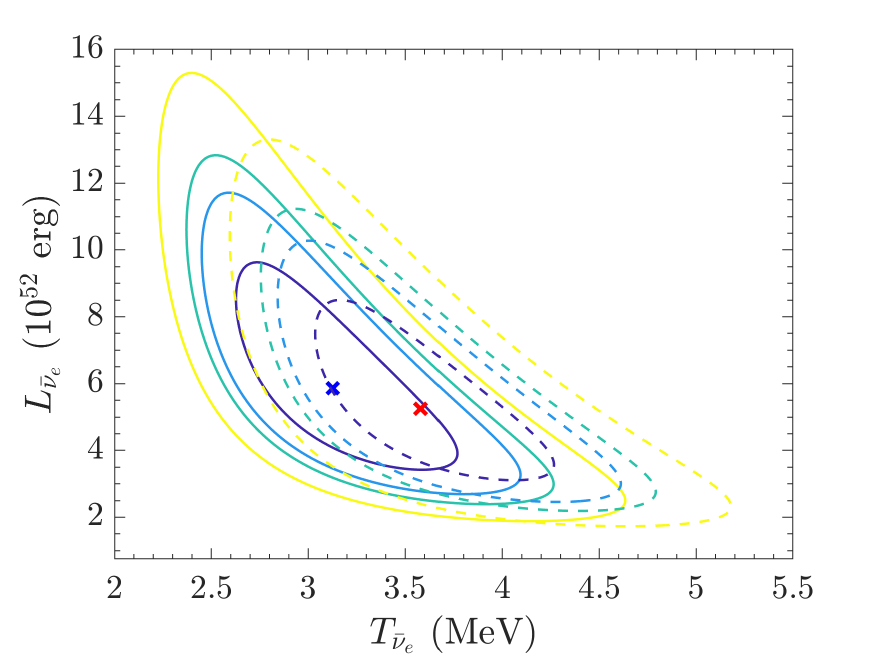}
\caption{Best fit and allowed regions for the $\bar{\nu}_e$ average energy and luminosity from 2D likelihood analysis of the 24 neutrino events from SN1987A observed by the Kamiokande-II, IMB and Baksan Collaborations. The results correspond to 10 $\%$ CL (yellow), 50 $\%$ CL (green), $90 \%$ (light blue) and 99 $\%$ (violet). The full lines are obtained taking the pinching parameter $\alpha = 2$, whereas dashed lines correspond to $\alpha = 3$. The results correspond to normal mass ordering. }
\label{fig:2Dcontours}
\end{center}
\end{figure}

Before presenting the results of the 7D likelihood analysis, we wish to give our 2D allowed regions for the $\bar{\nu}_e$ average energy, $\langle E_{\bar\nu_e}\rangle$, and luminosity, $L_{\bar\nu_e}$, in the absence of neutrino decay (Figure \ref{fig:2Dcontours}). For that, we use Eq.~\ref{eq:lh} considering $x = L_{\bar\nu_e},~\langle E_{\bar\nu_e}\rangle$. This analysis was made by several authors previously (see Refs.\cite{Vissani:2014doa,Lunardini:2004bj,Jegerlehner:1996kx}). 
 As one can see, the contours at 10 $\%$, 50 $\%$, $90 \%$ and 99 $\%$ agree well with those e.g. from \cite{Vissani:2014doa}. 

We explored the sensitivity to the pinching parameter $\alpha$ (Eq.\eqref{eq:PW}) that is usually kept fixed in this kind of analysis, because of the paucity of the events' ensemble. 
Figure \ref{fig:2Dcontours} shows the impact of the pinching parameter when $\alpha$ is varied from 2 to 3. 
The best-fit points for the $\bar{\nu}_e$ average energies and luminosities are $ 9.4  $ MeV and $5.9 \times 10^{52}$~erg, when $\alpha = 2$, and $  10.6  $ MeV and $5.2 \times 10^{52}$ erg
 for $\alpha = 3$. These results, valid for normal ordering, are obtained including Baksan data and the background. 
One can see that the pinching parameter modifies the allowed regions similarly
as aspects like the background, the detector's energy thresholds, or the implementation of Baksan events ({\it cfr} Figure 11 of Ref. \cite{Vissani:2014doa}). 

To check the quality of the fit, following Ref.~\cite{Vissani:2014doa}, we use the best-fit values of the luminosity and average energy, and we calculate the expected number of signal events at each of the experiments. Adding these numbers to the background events, we compute the Poisson probability of obtaining at least the number of events of Table~\ref{tab:SN1987Aevents}. For $\alpha = 2~(3)$, the probabilities are 75\% (75\%), 14\% (14\%), and 7\% (8\%) for Kamiokande-II, IMB, and Baksan, respectively. These results indicate the reliability of our best-fit values.
 
\subsection{The neutrino lifetime-to-mass ratio}

Let us now discuss the results when neutrinos are allowed to decay from the supernova surface to the Earth. 
As previously discussed, the exponential suppression of the flux gives a quantitative argument on a
characteristic $\tau/m$, depending on the source (Figure \ref{fig:taumtyp}). From the detection of $\bar{\nu}_e$ events in Kamiokande-II, IMB, and Baksan, one can exclude that neutrinos decayed completely. Since neutrinos traveled over a distance of 50 kpc, this translates to the lower bound $\tau/m \ge 10^{5} $ s/eV  \cite{Kamiokande-II:1987idp}.
This bound obviously holds without any assumption on the mass ordering (and considering invisible daughter particles or neutrinos). 
In a given mass ordering, if the constraint is applied to the lightest mass eigenstates, the other two 
could have a shorter lifetime-to-mass ratio.

This said, we turn to the results of the 7D unbinned likelihood analysis, which
exploits the spectral distortion due to neutrino decay. The seven parameters defining the likelihoods are the six parameters of the neutrino fluences for $\nu_e$, $\bar{\nu}_e$
and $\nu_x$\footnote{Note that the fluences for $\bar{\nu}_x$ are equal to $\nu_x$.} (Table~\ref{tab:priors}), and the lifetime-to-mass ratio. We fix the pinching parameters to $\alpha = 2.3$. We remind that in the case of NO and SH mass pattern and in the case of IO the neutrino and antineutrino sectors are connected
by the decays (Figure \ref{fig:decaypatterns}).
 Following  Ref.~\cite{Jegerlehner:1996kx}, in order to determine the profile likelihoods, we considered the average energies and luminosities
of $\nu_e$ and of $\bar{\nu}_x$ as ratios of $\bar{\nu}_e$. 
The priors used in the calculations are shown in Table \ref{tab:priors}.
\begin{table}[]
\begin{tabular}{c|c|lc|cc|}
\cline{2-2} \cline{5-6}
\multicolumn{1}{l|}{}                                                                                 & $\bar\nu_e$ &                       &             & $\nu_e$   & $\nu_x$   \\ \cline{1-2} \cline{4-6} 
\multicolumn{1}{|c|}{\begin{tabular}[c]{c}$L_{\nu_{\alpha}}$\\ {[}$10^{52}$ erg{]}\end{tabular}}       & [1, 13]      & \multicolumn{1}{l|}{} & $t_{\nu_{\alpha}}$ & [0.5, 1.8] & [0.5, 1.8] \\ \cline{1-2} \cline{4-6} 
\multicolumn{1}{|c|}{\begin{tabular}[c]{c}$\langle E_{\nu_{\alpha}} \rangle$\\ {[}MeV{]}\end{tabular}} & [7, 17.5]    & \multicolumn{1}{l|}{} & $k_{\nu_{\alpha}}$ & [0.8, 1.8] & [0.8, 1.8] \\ \cline{1-2} \cline{4-6} 
\end{tabular}
\caption{Parameters considered in the 7D likelihood analysis and priors inside which they are free to vary. The quantities $ t_{\nu_\alpha}$ and  $k_{\nu_\alpha} $ are the
ratio of the luminosities $t_{\nu_\alpha} = L_{\nu_\alpha}/L_{\bar\nu_e}$ and of the average energies $k_{\nu_\alpha} = \langle E_{\nu_\alpha} \rangle / \langle E_{\bar\nu_e} \rangle$ for $\nu_{\alpha} = \nu_e$ (second column) and $\nu_x = \nu_{\mu}, \nu_{\tau}, \bar{\nu}_{\mu}$ and $\bar{\nu}_{\tau}$ (third column).}\label{tab:priors}
\end{table}

To include decay, we considered a full 3$\nu$ flavor framework and solved Eqs. \eqref{eq:transf}, including the MSW effect (Eq.~\eqref{eq:MSWnubar}) 
to determine the neutrino fluences at the supernova surface, before neutrinos decay. 
Both the NO and IO cases are considered. For the former, either a SH or a QD mass pattern is assumed.
Note that a QD mass pattern is already in tension with upper bounds on the sum of neutrino masses from cosmological observations \cite{ParticleDataGroup:2022pth}.

\begin{figure}
\begin{center}
\includegraphics[scale=0.4]{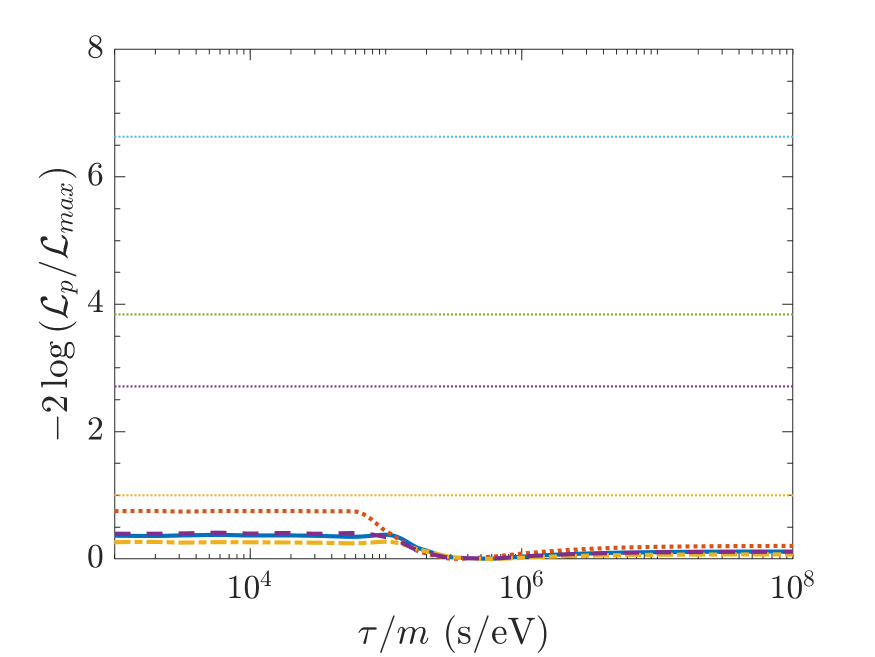}
\caption{Normal mass ordering and SH mass pattern: profile likelihood ratios from the 7D likelihood analysis of SN1987A events. 
The calculations of the neutrino fluxes include flavor conversion due to the MSW effect and neutrino decay in a 3$\nu$ flavor framework. 
The curves correspond to the cases without Baksan events and with background (full line),
with Baksan events and background (dot-dashed line), without Baksan events and without background (dotted)
and with Baksan events and without background (dashed line). The dashed straight lines from bottom to top correspond
to 68 $\%$, 90 $\%$, 95 $\%$, and 99 $\%$ CL.   }
\label{fig:NOSH}
\end{center}
\end{figure}

\begin{figure}
\begin{center}
\includegraphics[scale=0.4]{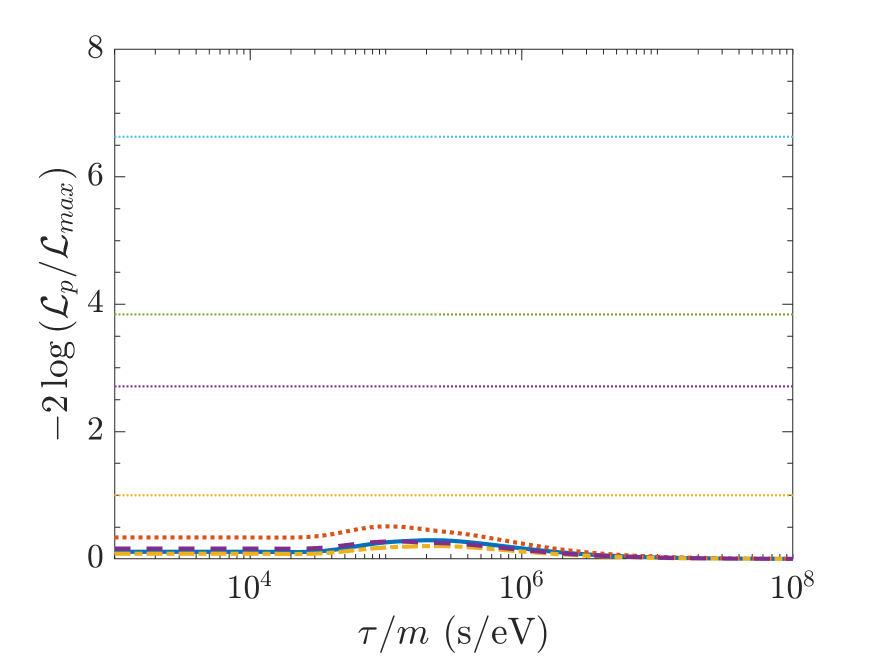}
\caption{Normal mass ordering and QD mass pattern: profile likelihood ratios from the 7D likelihood analysis of SN1987A events. 
The calculations of the neutrino fluxes include flavor conversion due to the MSW effect and neutrino decay in a 3$\nu$ flavor framework. 
The curves correspond to the cases without Baksan events and with background (full line),
with Baksan events and background (dot-dashed line), without Baksan events and without background (dotted)
and with Baksan events and without background (dashed line). The dashed straight lines from bottom to top correspond
to 68 $\%$, 90 $\%$, 95 $\%$, and 99 $\%$ CL.   }
\label{fig:NOQD}
\end{center}
\end{figure}

Figures \ref{fig:NOSH} and  \ref{fig:NOQD} show the results of the likelihood analysis for NO for $\tau/m \in [10^{3}, 10^{8}]$ s/eV.
One can see that for NO, both for a SH and for a QD mass pattern the profile likelihood ratios are practically flat showing little
sensitivity to neutrino decay. This is due to the fact that for these cases the spectral distortions are too small to
produce significant variations in the SN1987A events. 
This conclusion holds even when adding Baksan events or without the background. 

\begin{figure}
\begin{center}
\includegraphics[scale=0.4]{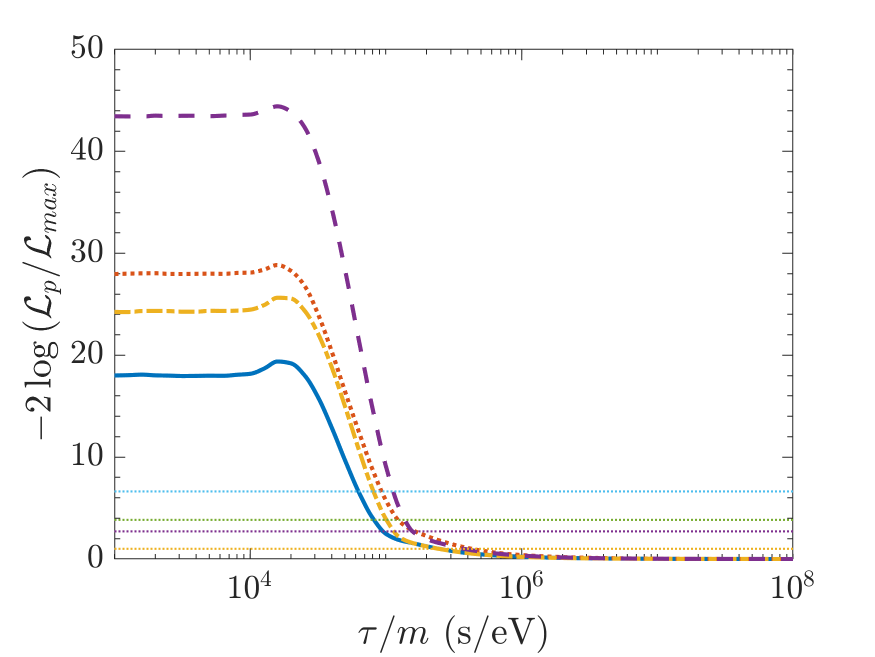}
\caption{Inverted mass ordering: profile likelihood ratios from the 7D likelihood analysis of SN1987A events. 
The calculations of the neutrino fluxes include flavor conversion due to the MSW effect and neutrino decay in a 3$\nu$ flavor framework. 
The curves correspond to the cases without Baksan events and with background (full line),
with Baksan events and background (dot-dashed line), without Baksan events and without background (dotted)
and with Baksan events and without background (dashed line). The dashed straight lines from bottom to top correspond
to 68 $\%$, 90 $\%$, 95 $\%$, and 99 $\%$ CL.   }
\label{fig:IO}
\end{center}
\end{figure}

On the contrary, for the case of IO, the spectral modification due to neutrino decay is such that the results of the profile likelihood ratios
show a sensitivity to $\tau/m$ (Figure \ref{fig:IO}). Note that a strong impact of neutrino decay on the DSNB was also found in Refs.\cite{Fogli:2004gy,Ivanez-Ballesteros:2022szu} for IO. From the numerical results we obtain the
lower bound 
\begin{equation}\label{eq:IOtauoverm}
{\tau \over m} \ge 2.4 \times 10^{5}~ {\rm s/eV} ~~(1.2 \times 10^{5})~ {\rm s/eV} ~~~~~68 \%~ (90 \%) ~ {\rm CL} \ , 
\end{equation}
for the $\nu_2$ and the $\nu_1$ mass eigenstates, when Baksan events are included and with the background. The influence of Baksan events is generally small, whereas
the one of the background is larger. 
Indeed, the lower bound Eq.\eqref{eq:IOtauoverm} increases to $3.8 \times 10^{5}~ {\rm s/eV}$ (68 $\%$ CL) when Baksan events are excluded, 
and to $4.3 \times 10^{5}~ {\rm s/eV} $ (68 $\%$ CL) if the background is excluded as well.
Similarly,  the bound \eqref{eq:IOtauoverm} varies from $1.2 \times 10^{5}~ {\rm s/eV}$ to $1.6 \times 10^{5}~ {\rm s/eV}$ ($90 \%$ CL) when Baksan events are excluded, and to 
$1.7 \times 10^{5}~ {\rm s/eV}$  ($90 \%$ CL) without the background. We checked that these results do not vary significantly when
considering values of $\alpha$ in the range between 2 and 3. This range of values describes well the fluences obtained by supernova simulations.

\subsection{Discussion}

\begin{table*}[t]
\centering
\begin{tabular}{|c r|c|c|c|}
\hline 
Analysis & Ref. & Mass ordering & Decaying $\nu$ & Lower limits [s/eV] \\ \hline
Atmospheric and LBL & \cite{Gonzalez-Garcia:2008mgl} & NO & $\nu_3$ & $9.3 \times 10^{-11}$ \\ \hline
Reactor & \cite{Porto-Silva:2020gma} &NO (IO) & $\nu_3$ ($\nu_2$) & $0.1$ ($1.4$) $ \times 10^{-9}$ \\ \hline
\multirow{5}{*}{Solar}  & \cite{Berryman:2014qha} & independent & \begin{tabular}[c]{c} $\nu_1$ \\ $\nu_2$ \end{tabular}& \begin{tabular}[c]{c} $4.1 \times 10^{-3}$\\ $7.1 \times 10^{-4}$ \end{tabular}\\ \cline{3-5}
  & \cite{SNO:2018pvg} & independent & $\nu_2$ & $1.92 \times 10^{-3}$  \\ \cline{3-5}
  & \cite{Funcke:2019grs} & NO, SH & \begin{tabular}[c]{c} $\nu_2$ \\ $\nu_3$ \end{tabular}& \begin{tabular}[c]{c} $1.1 \times 10^{-3}$\\ $2.2 \times 10^{-5}$ \end{tabular} \\ \cline{3-5}
  & \cite{Funcke:2019grs} & NO, QD & \begin{tabular}[c]{c} $\nu_2$ \\ $\nu_3$ \end{tabular}& \begin{tabular}[c]{c} $6.7 \times 10^{-4}$\\ $1.3 \times 10^{-4}$ \end{tabular} \\ \cline{3-5}
  & \cite{Picoreti:2021yct} & NO, SH & \begin{tabular}[c]{c} $\nu_3 \to \nu_1$ \\ $\nu_3 \to \nu_2$ \end{tabular}& \begin{tabular}[c]{c} $7 \times 10^{-5}$\\ $1 \times 10^{-5}$ \end{tabular} \\ \hline
Ultra-high energy $\nu$ & \cite{Pagliaroli:2015rca} & NO (IO) & $\nu_2$, $\nu_3$ ($\nu_1$, $\nu_2$) & $1.0 \times 10^{3}$ \\ \hline \hline
SN1987A (This work)&  & IO & $\nu_1$, $\nu_2$ & $1.2 \times 10^{5}$ \\ \hline
\end{tabular}
\caption{Lower limits to the neutrino lifetime-to-mass ratio from different studies including the result of this work (last row). All the results are given at 90\% CL except 
for the atmospheric and LBL one at 99 \% CL, for the first solar limit and the astrophysical limit which are given at 95\% CL. The first two columns indicate the source of neutrinos and the reference to the study. The third and fourth columns show the mass ordering assumed in the analysis and the decaying neutrino. Finally, the last column presents the lower limits of $\tau/m$.}
\label{tab:limits}
\end{table*}

As mentioned in the introduction, several studies have constrained neutrino nonradiative decay through the observation of neutrinos from different sources. Table~\ref{tab:limits} summarizes the limits obtained by different studies. It is important to notice that most of these limits are obtained for normal mass ordering and there is a limited number of studies that consider inverted mass ordering. We observe from Table~\ref{tab:limits} that the least restrictive limits are obtained from the combined analysis of atmospheric and accelerator data~\cite{Gonzalez-Garcia:2008mgl} for NO and from reactor data for IO~\cite{Porto-Silva:2020gma}. From solar data, the limits obtained by different analyses span from $1.0\times 10^{-5}$ to $1.92\times 10^{-3}$~s/eV. Astrophysical neutrinos coming from further sources give tighter limits to the lifetime-to-mass ratio. In particular, IceCube data disfavors neutrino decay at 95\% CL for $\tau /m < 10^3$~s/eV~\cite{Pagliaroli:2015rca}.

Even tighter constraints can be obtained by studying the effect of neutrino decay on the CMB. This process can produce a loss of anisotropic stress which is incompatible with CMB observations. The study made in Ref.~\cite{Chen:2022idm} considered different decay scenarios. For the case of two decay channels and inverted mass ordering, they obtained their tightest constraint on the neutrino lifetime: $(5 \to 6) \times 10^7$~s for the lightest neutrino mass between $10^{-3}$ and $10^{-1}$~eV. For normal mass ordering, in the same mass range, they found $ \tau > (2 \to 3) \times 10^7$~s.

Ref.~\cite{Frieman:1987as} studied this decay in relation to the events detected at Kamiokande-II. However, they did not perform a detailed statistical analysis and did not derive any bounds on the neutrino lifetime. Studies like Refs.~\cite{Kachelriess:2000qc, Farzan:2002wx} considered a Majoron model in matter and they obtained limits on the neutrino-Majoron couplings. 

Our analysis of SN1987A data improves with respect to previous studies Refs.~\cite{Kachelriess:2000qc, Frieman:1987as} in several aspects. On one hand, as opposed to Ref.~\cite{Frieman:1987as}, we performed a detailed statistical analysis of the data and we included the data of the three detectors. With respect to Refs.~\cite{Kachelriess:2000qc, Farzan:2002wx}, the main difference comes from the fact that we study decay in vacuum while they consider neutrino-Majoron interactions in matter. Note that Ref.~\cite{Kachelriess:2000qc} considers only diagonal couplings\footnote{The Lagrangian considered in Ref.~\cite{Kachelriess:2000qc} is different from the one we presented in Eq.~\ref{eq:lagrangian}.} and sets $g_{33} = 0$ which is effectively a two-flavor approximation. Their results are only valid for normal ordering.

Regarding our results, in the case of normal ordering, we have seen that the events from SN1987A have very little sensitivity to neutrino decay. Therefore, our results do not improve or reject any previous bounds. On the contrary, in the case of inverted ordering, we have obtained lower limits to the lifetime-to-mass ratio (Eq.(\ref{eq:IOtauoverm})). These limits are tighter with respect to the ones shown in Table~\ref{tab:limits}. After the model-dependent limits obtained from cosmological constraints, our results are the tightest limits on $\tau/m$ in the case of IO.

\section{Conclusions}
In conclusion, we have performed a detailed analysis of the 24 neutrino events from SN1987A,
detected by Kamiokande-II, IMB, and Baksan detectors, in the presence of neutrino non-radiative two-body decay in a full 3$\nu$ framework.
To this aim, we have used the solutions of the transfer equations,
considered the two possible neutrino mass ordering and the still allowed mass patterns. 

When neutrino decay is excluded, our 2D likelihood analysis conveys allowed regions for the $\bar{\nu}_e$
average energy and luminosity in agreement with previous works. The results show that if the pinching
parameter is allowed to vary, the allowed regions shift similarly to the uncertainties usually considered in this kind of analysis. 

As for the 7D likelihood investigation, the inclusion of neutrino decay produces spectral distortions 
and influences the predicted events for inverse beta-decay in the three detectors. In the case of normal ordering, we found that modifications are too small to yield significant
bounds on the lifetime-to-mass ratio. 
It is to be noted that numerous analyses on neutrino decay existing in the literature assume normal ordering, and often a strongly hierarchical mass pattern. Therefore our results for normal ordering do not represent an improvement with respect to either the bounds using other neutrino sources or the bounds obtained by the argument that, since $\bar\nu_e$ from SN1987A were actually seen in the detectors, complete decay is rejected.

On the contrary, in inverted mass ordering, our 7D likelihood analysis of the SN1987A events
gives the interesting lower bound \eqref{eq:IOtauoverm} on $\tau/m$. Our findings provide
one of the tightest bounds on the lifetime-to-mass ratio, for inverted mass ordering, for the $\nu_2$ and the $\nu_1$ mass eigenstates.

\section*{Acknowledgments}
M.~Cristina Volpe thanks Francesco Vissani for useful discussions and
for pointing out Ref. \cite{Pagliaroli:2015rca} after completion of this work. 

\appendix
\section{Neutrino fluxes when neutrinos decay}
We assume that the lightest (anti)neutrino $\nu_l$ is stable and that the heaviest one, $\nu_h$, and the intermediate mass eigenstate $\nu_i$ can decay into the lighter one. 
Moreover, we do not consider the possibility that the neutrinos decay in the supernova, i.e. $\Gamma R_{\rm SN} \ll 1$. By integrating the transfer equations \eqref{eq:transf} one has the following expressions for the antineutrino fluxes \cite{Ando:2004qe} 

\begin{equation}
\label{eq:flux_h}
    F_{\bar\nu_h} (E_{\nu}, L) = e^{-\Gamma_h (E_{\nu})  L} ~F_{\bar\nu_h} (E_{\nu}, R_{\rm SN})   ,
\end{equation}
with $\Gamma_h (E_{\nu}) = \Gamma_{hi} (E_{\nu}) + \Gamma_{hl} (E_{\nu})  $ being the decay rate in the laboratory frame,
$R_{\rm SN}$ the supernova surface and $L$ the distance between SN1987A and the Earth.
 
For the intermediate decaying eigenstate the following expression holds
\begin{equation}
\label{eq:flux_i}
\begin{split}
    F_{\bar\nu_i} (E_{\nu}, L) &= e^{-\Gamma_i (E_{\nu}) L} ~ F_{\bar\nu_i} (E_{\nu}, R_{\rm SN})  \\ & + \int_{E_{\nu}}^\infty dE_{\nu}' \frac{e^{-\Gamma_i (E_{\nu}) L} - e^{-\Gamma_h (E_{\nu}') L}}{\Gamma_h (E_{\nu}') - \Gamma_i (E_{\nu})} J_{h\rightarrow i}(E_{\nu}',E_{\nu}) ,
\end{split}
\end{equation}
where $\Gamma_i (E_{\nu}) = \Gamma_{il} (E_{\nu})$; whereas for the light decaying eigenstate one has

\begin{equation}
\label{eq:flux_l}
\begin{split}
    F_{\bar\nu_l} (E_{\nu}, L) &= F_{\bar\nu_l} (E_{\nu}, R_{\rm SN}) 
    \\ & + \int_{E_{\nu}}^\infty dE_{\nu}' \Bigg[ \Bigg. \frac{1 - e^{-\Gamma_h (E_{\nu}') L}}{\Gamma_h (E_{\nu}')} J_{h\rightarrow l}(E_{\nu}',E_{\nu}) 
    \\ & +  \frac{1 - e^{-\Gamma_i (E_{\nu}') L}}{\Gamma_i (E_{\nu}')} J_{i\rightarrow l}(E_{\nu}',E_{\nu}) 
    \\ & +  \int_{E_{\nu}'}^\infty dE_{\nu}'' \Big( \Big.
    \frac{1 - e^{-\Gamma_i (E_{\nu}') L}}{\Gamma_i (E_{\nu}')} 
    \\ &  -\frac{1 - e^{-\Gamma_h (E_{\nu}'') L}}{\Gamma_h (E_{\nu}'')} \Big.\Big) J_{h\rightarrow i \rightarrow l} (E_{\nu}'',E_{\nu}',E_{\nu})
        \Bigg. \Bigg],
\end{split}
\end{equation}
and similarly for the neutrino fluxes. 
In these expressions, we define

\begin{equation}
    \begin{split}
    J_{h \rightarrow i \rightarrow l}(E_{\nu}'',E_{\nu}',E_{\nu})  & = \frac{1}{\Gamma_h (E_{\nu}'') - \Gamma_i (E_{\nu}')} 
    \\ 
   & \times \left[ 
     \psi_{\rm h.c.}(E_{\nu}',E_{\nu}) \Gamma_{\bar{\nu}_i \rightarrow \bar{\nu}_l} (E_{\nu}')  J_{h \rightarrow i} (E_{\nu}'',E_{\nu}')  
    \right. 
    \\ & + \left.  \psi_{\rm h.f.}(E_{\nu}',E_{\nu})  \Gamma_{{\nu}_i \rightarrow \bar{\nu}_l} (E_{\nu}') \tilde{J}_{h \rightarrow i} (E_{\nu}'',E_{\nu}') 
    \right]
    \end{split}
\end{equation}

\begin{equation}
\begin{split}
    J_{h \rightarrow i}(E_{\nu}',E_{\nu}) &= \psi_{\rm h.c.}(E_{\nu}',E_{\nu}) \Gamma_{\bar{\nu}_h \rightarrow \bar{\nu}_i} (E_{\nu}') F_{\bar{\nu}_h}(E_{\nu}', R_{\rm SN})
    \\ & + \psi_{\rm h.f.}(E_{\nu}',E_{\nu}) \Gamma_{\nu_h \rightarrow \bar{\nu}_i} (E_{\nu}') F_{\nu_h}(E_{\nu}', R_{\rm SN}) ,
\end{split}
\end{equation}

\begin{equation}
\begin{split}
    \tilde{J}_{h \rightarrow i}(E_{\nu}',E_{\nu}) &= \psi_{\rm h.c.}(E_{\nu}',E_{\nu}) \Gamma_{\nu_h \rightarrow \nu_i} (E_{\nu}') F_{\nu_h}(E_{\nu}', R_{\rm SN})
    \\ &+\psi_{\rm h.f.}(E_{\nu}',E_{\nu})  \Gamma_{\bar\nu_h \rightarrow \nu_i} (E_{\nu}')  F_{\bar{\nu}_h}(E_{\nu}',R_{\rm SN}) .
\end{split}
\end{equation}

The expressions for $\psi_{\rm h.c.}$ and $\psi_{\rm h.f.}$ are given in the main text for the different mass patterns considered.

\section{SN1987A neutrino events}
For completeness, we give in Table~\ref{tab:SN1987Aevents} the information used in our analysis, concerning the 24 neutrino events recorded by Kamiokande-II \cite{Kamiokande-II:1987idp}, IMB \cite{Bionta:1987qt} and Baksan \cite{Alekseev:1988gp} detectors when SN1987A occurred. 

\begin{table*}[]
\begin{center}
\begin{tabular}{|lrr|c|lrcc|}
\hline
\multicolumn{1}{|c}{} & \multicolumn{1}{c}{\begin{tabular}[c]{c}Time\\ {[}ms{]}\end{tabular}} & \multicolumn{1}{c|}{\begin{tabular}[c]{c}Energy\\ {[}MeV{]}\end{tabular}} & \begin{tabular}[c]{c}Background\\ {[}Hz/MeV{]}\end{tabular} & \multicolumn{1}{c}{} & \multicolumn{1}{c}{\begin{tabular}[c]{c}Time\\ {[}ms{]}\end{tabular}} & \multicolumn{1}{c|}{\begin{tabular}[c]{c}Energy\\ {[}MeV{]}\end{tabular}} & \begin{tabular}[c]{c}Background\\ {[}Hz/MeV{]}\end{tabular} \\ \hline
                      &                                                                             &                                                                                 &                                                                   & I1                   & 0                                                                           & \multicolumn{1}{c|}{38 $\pm$ 7}                                                 & $10^{-5}$   ?                                                     \\
K1                    & 0                                                                           & 20.0 $\pm$ 2.9                                                                  & $10^{-5}$                                                         & I2                   & 412                                                                         & \multicolumn{1}{c|}{37 $\pm$ 7}                                                 & $10^{-5}$ ?                                                       \\
K2                    & 107                                                                         & 13.5 $\pm$ 3.2                                                                  & $5.4\times10^{-4}$                                                & I3                   & 650                                                                         & \multicolumn{1}{c|}{28 $\pm$ 6}                                                 & $10^{-5}$ ?                                                       \\
K3                    & 303                                                                         & 7.5 $\pm$ 2.0                                                                   & $2.4\times10^{-2}$                                                & I4                   & 1141                                                                        & \multicolumn{1}{c|}{39 $\pm$ 7}                                                 & $10^{-5}$ ?                                                       \\
K4                    & 324                                                                         & 9.2 $\pm$ 2.7                                                                   & $2.8\times10^{-3}$                                                & I5                   & 1562                                                                        & \multicolumn{1}{c|}{36 $\pm$ 9}                                                 & $10^{-5}$ ?                                                       \\
K5                    & 507                                                                         & 12.8 $\pm$ 2.9                                                                  & $5.3\times10^{-4}$                                                & I6                   & 2684                                                                        & \multicolumn{1}{c|}{36 $\pm$ 6}                                                 & $10^{-5}$ ?                                                       \\
K6                    & 1541                                                                        & 35.4 $\pm$ 8.0                                                                  & $5.0\times10^{-6}$                                                & I7                   & 5010                                                                        & \multicolumn{1}{c|}{19 $\pm$ 5}                                                 & $10^{-5}$ ?                                                       \\
K7                    & 1728                                                                        & 21.0 $\pm$ 4.2                                                                  & $10^{-5}$                                                         & I8                   & 5582                                                                        & \multicolumn{1}{c|}{22 $\pm$ 5}                                                 & $10^{-5}$ ?                                                       \\
K8                    & 1915                                                                        & 19.8 $\pm$ 3.2                                                                  & $10^{-5}$                                                         &                      &                                                                             & \multicolumn{1}{c|}{}                                                           &                                                                   \\
K9                    & 9219                                                                        & 8.6 $\pm$ 2.7                                                                   & $4.2\times10^{-3}$                                                & B1                   & 0                                                                           & \multicolumn{1}{c|}{12.0 $\pm$ 2.4}                                             & $8.4\times10^{-4}$                                                \\
K10                   & 10433                                                                       & 13.0 $\pm$ 2.6                                                                  & $4.0\times10^{-4}$                                                & B2                   & 435                                                                         & \multicolumn{1}{c|}{17.9 $\pm$ 3.6}                                             & $1.3\times10^{-3}$                                                \\
K11                   & 12439                                                                       & 8.9 $\pm$ 2.9                                                                   & $3.2\times10^{-3}$                                                & B3                   & 1710                                                                        & \multicolumn{1}{c|}{23.5 $\pm$ 4.7}                                             & $1.2\times10^{-3}$                                                \\
                      &                                                                             &                                                                                 &                                                                   & B4                   & 7687                                                                        & \multicolumn{1}{c|}{17.5 $\pm$ 3.5}                                             & $1.3\times10^{-3}$                                                \\
                      & \multicolumn{1}{l}{}                                                        & \multicolumn{1}{l|}{}                                                           &                                                                   & B5                   & 9099                                                                        & 20.3 $\pm$ 4.1                                                                  & $1.3\times10^{-3}$                                                \\ \hline
\end{tabular}

\caption{SN1987A events: time (from first detection in each detector) and energy of the neutrino events seen in Kamiokande-II (K1-K11), in IMB (I1-I8) and in Baksan (B1-B5) detectors. The related backgrounds are also shown \cite{Vissani:2014doa}. The background rate of IMB is unknown, however, we use the values represented in the table with a question mark. Note that time information in Kamiokande-II has an uncertainty of $\pm$1 min, whereas the ones of IMB and Baksan of (-54 s, +2 s) and $\pm 50 $ ms respectively. Baksan data point to a $\bar{\nu}_e$ luminosity larger than the one from Kamiokande-II (by a factor of about 3) \cite{Alekseev:1988gp}. Note also that the time spread of the events in Kamiokande-II and Baksan show a double structure (separated by 7 s) whereas IMB data are compatible with an exponential \cite{Alekseev:1988gp}. (The double bump feature is likely an accidental feature.)} \label{tab:SN1987Aevents}
\end{center}
\end{table*}


\begin{thebibliography}{00}



\bibitem{Kamiokande-II:1987idp}
K.~Hirata \textit{et al.} [Kamiokande-II],
Phys. Rev. Lett. \textbf{58}, 1490-1493 (1987).

\bibitem{Bionta:1987qt}
R.~M.~Bionta, G.~Blewitt, C.~B.~Bratton, D.~Casper, A.~Ciocio, R.~Claus, B.~Cortez, M.~Crouch, S.~T.~Dye and S.~Errede, \textit{et al.}
Phys. Rev. Lett. \textbf{58}, 1494 (1987).

\bibitem{Alekseev:1988gp}
E.~N.~Alekseev, L.~N.~Alekseeva, I.~V.~Krivosheina and V.~I.~Volchenko,
Phys. Lett. B \textbf{205}, 209-214 (1988).

\bibitem{Aglietta:1987it}
M.~Aglietta, G.~Badino, G.~Bologna, C.~Castagnoli, A.~Castellina, W.~Fulgione, P.~Galeotti, O.~Saavedra, G.~Trinchero and S.~Vernetto, \textit{et al.}
EPL \textbf{3}, 1315-1320 (1987).

\bibitem{Super-Kamiokande:1998kpq}
Y.~Fukuda \textit{et al.} [Super-Kamiokande],
Phys. Rev. Lett. \textbf{81}, 1562-1567 (1998)
[arXiv:hep-ex/9807003 [hep-ex]].

\bibitem{SNO:2001kpb}
Q.~R.~Ahmad \textit{et al.} [SNO],
Phys. Rev. Lett. \textbf{87}, 071301 (2001),
[arXiv:nucl-ex/0106015 [nucl-ex]].

\bibitem{KamLAND:2002uet}
K.~Eguchi \textit{et al.} [KamLAND],
Phys. Rev. Lett. \textbf{90}, 021802 (2003),
[arXiv:hep-ex/0212021 [hep-ex]].

\bibitem{Volpe:2023met}
M.~C.~Volpe,
[arXiv:2301.11814 [hep-ph]].

\bibitem{Duan:2010bg}
H.~Duan, G.~M.~Fuller and Y.~Z.~Qian,
Ann. Rev. Nucl. Part. Sci. \textbf{60} (2010), 569-594
[arXiv:1001.2799 [hep-ph]].

\bibitem{Mirizzi:2015eza}
A.~Mirizzi, I.~Tamborra, H.~T.~Janka, N.~Saviano, K.~Scholberg, R.~Bollig, L.~Hudepohl and S.~Chakraborty,
Riv. Nuovo Cim. \textbf{39} (2016) no.1-2, 1-112
[arXiv:1508.00785 [astro-ph.HE]].

\bibitem{Tamborra:2020cul}
I.~Tamborra and S.~Shalgar,
Ann. Rev. Nucl. Part. Sci. \textbf{71} (2021), 165-188
[arXiv:2011.01948 [astro-ph.HE]].

\bibitem{ParticleDataGroup:2022pth}
R.~L.~Workman \textit{et al.} [Particle Data Group],
PTEP \textbf{2022}, 083C01 (2022).


\bibitem{Capozzi:2021fjo}
F.~Capozzi, E.~Di Valentino, E.~Lisi, A.~Marrone, A.~Melchiorri and A.~Palazzo,
Phys. Rev. D \textbf{104}, no.8, 083031 (2021),
[arXiv:2107.00532 [hep-ph]].

\bibitem{Gonzalez-Garcia:2014bfa}
M.~C.~Gonzalez-Garcia, M.~Maltoni and T.~Schwetz,
JHEP \textbf{11} (2014), 052
[arXiv:1409.5439 [hep-ph]].

\bibitem{Esteban:2018azc}
I.~Esteban, M.~C.~Gonzalez-Garcia, A.~Hernandez-Cabezudo, M.~Maltoni and T.~Schwetz,
JHEP \textbf{01} (2019), 106
[arXiv:1811.05487 [hep-ph]].


\bibitem{KATRIN:2021uub}
M.~Aker \textit{et al.} [KATRIN],
Nature Phys. \textbf{18}, no.2, 160-166 (2022),
[arXiv:2105.08533 [hep-ex]].

\bibitem{Planck:2018vyg}
N.~Aghanim \textit{et al.} [Planck],
Astron. Astrophys. \textbf{641} (2020), A6
[erratum: Astron. Astrophys. \textbf{652} (2021), C4]
[arXiv:1807.06209 [astro-ph.CO]].

\bibitem{Giunti:2014ixa}
C.~Giunti and A.~Studenikin,
Rev. Mod. Phys. \textbf{87}, 531 (2015),
[arXiv:1403.6344 [hep-ph]].



\bibitem{Bahcall:1972my}
J.~N.~Bahcall, N.~Cabibbo and A.~Yahil,
Phys. Rev. Lett. \textbf{28}, 316-318 (1972).

\bibitem{Gonzalez-Garcia:2008mgl}
M.~C.~Gonzalez-Garcia and M.~Maltoni,
Phys. Lett. B \textbf{663}, 405-409 (2008),
[arXiv:0802.3699 [hep-ph]].

\bibitem{Beacom:2002cb}
J.~F.~Beacom and N.~F.~Bell,
Phys. Rev. D \textbf{65}, 113009 (2002),
[arXiv:hep-ph/0204111 [hep-ph]].



\bibitem{SNO:2018pvg}
B.~Aharmim \textit{et al.} [SNO],
Phys. Rev. D \textbf{99}, no.3, 032013 (2019)
[arXiv:1812.01088 [hep-ex]].

\bibitem{Barenboim:2020vrr}
G.~Barenboim, J.~Z.~Chen, S.~Hannestad, I.~M.~Oldengott, T.~Tram and Y.~Y.~Y.~Wong,
JCAP \textbf{03}, 087 (2021),
[arXiv:2011.01502 [astro-ph.CO]].

\bibitem{Chen:2022idm}
J.~Z.~Chen, I.~M.~Oldengott, G.~Pierobon and Y.~Y.~Y.~Wong,
Eur. Phys. J. C \textbf{82}, no.7, 640 (2022),
[arXiv:2203.09075 [hep-ph]].

\bibitem{Beacom:2002vi}
J.~F.~Beacom, N.~F.~Bell, D.~Hooper, S.~Pakvasa and T.~J.~Weiler,
Phys. Rev. Lett. \textbf{90}, 181301 (2003)
[arXiv:hep-ph/0211305 [hep-ph]].

\bibitem{IceCube:2013low}
M.~G.~Aartsen \textit{et al.} [IceCube],
Science \textbf{342}, 1242856 (2013),
[arXiv:1311.5238 [astro-ph.HE]].

\bibitem{Pagliaroli:2015rca}
G.~Pagliaroli, A.~Palladino, F.~L.~Villante and F.~Vissani,
Phys. Rev. D \textbf{92}, no.11, 113008 (2015),
[arXiv:1506.02624 [hep-ph]].

\bibitem{Denton:2018aml}
P.~B.~Denton and I.~Tamborra,
Phys. Rev. Lett. \textbf{121} (2018) no.12, 121802
[arXiv:1805.05950 [hep-ph]].

\bibitem{Abdullahi:2020rge}
A.~Abdullahi and P.~B.~Denton,
Phys. Rev. D \textbf{102} (2020) no.2, 023018
[arXiv:2005.07200 [hep-ph]].


\bibitem{Fogli:2004gy}
G.~L.~Fogli, E.~Lisi, A.~Mirizzi and D.~Montanino,
Phys. Rev. D \textbf{70}, 013001 (2004),
[arXiv:hep-ph/0401227 [hep-ph]].

\bibitem{Tabrizi:2020vmo}
Z.~Tabrizi and S.~Horiuchi,
JCAP \textbf{05}, 011 (2021),
[arXiv:2011.10933 [hep-ph]].


\bibitem{DeGouvea:2020ang}
A.~De Gouv\^ea, I.~Martinez-Soler, Y.~F.~Perez-Gonzalez and M.~Sen,
Phys. Rev. D \textbf{102}, 123012 (2020),
[arXiv:2007.13748 [hep-ph]].

\bibitem{Ivanez-Ballesteros:2022szu}
P.~Ivanez-Ballesteros and M.~C.~Volpe,
Phys. Rev. D \textbf{107}, no.2, 023017 (2023),
[arXiv:2209.12465 [hep-ph]].

\bibitem{deGouvea:2019goq}
A.~de Gouvea, I.~Martinez-Soler and M.~Sen,
Phys. Rev. D \textbf{101}, no.4, 043013 (2020),
[arXiv:1910.01127 [hep-ph]].

\bibitem{Raffelt:1996wa}
G.~G.~Raffelt,
{\it ``Stars as laboratories for fundamental physics: The astrophysics of neutrinos, axions, and other weakly interacting particles,'}
1996, The University of Chicago Press.

\bibitem{Fiorillo:2022cdq}
D.~F.~G.~Fiorillo, G.~G.~Raffelt and E.~Vitagliano,
Phys. Rev. Lett. \textbf{131} (2023) no.2, 021001
[arXiv:2209.11773 [hep-ph]].

\bibitem{Kachelriess:2000qc}
M.~Kachelriess, R.~Tomas and J.~W.~F.~Valle,
Phys. Rev. D \textbf{62}, 023004 (2000),
[arXiv:hep-ph/0001039 [hep-ph]].

\bibitem{Farzan:2002wx}
Y.~Farzan,
Phys. Rev. D \textbf{67}, 073015 (2003)
[arXiv:hep-ph/0211375 [hep-ph]].


\bibitem{Frieman:1987as}
J.~A.~Frieman, H.~E.~Haber and K.~Freese,
Phys. Lett. B \textbf{200}, 115-121 (1988).

\bibitem{Keil:2002in}
M.~T.~Keil, G.~G.~Raffelt and H.~T.~Janka,
Astrophys. J. \textbf{590}, 971-991 (2003),
[arXiv:astro-ph/0208035 [astro-ph]].


\bibitem{Wolfenstein:1977ue}
L.~Wolfenstein,
Phys. Rev. D \textbf{17}, 2369-2374 (1978).

\bibitem{Mikheev:1986wj}
S.~P.~Mikheev and A.~Y.~Smirnov,
Nuovo Cim. C \textbf{9}, 17-26 (1986).




\bibitem{Dighe:1999bi}
A.~S.~Dighe and A.~Y.~Smirnov,
Phys. Rev. D \textbf{62}, 033007 (2000),
[arXiv:hep-ph/9907423 [hep-ph]].


\bibitem{Berryman:2014qha}
J.~M.~Berryman, A.~de Gouvea and D.~Hernandez,
Phys. Rev. D \textbf{92} (2015) no.7, 073003
doi:10.1103/PhysRevD.92.073003
[arXiv:1411.0308 [hep-ph]].

\bibitem{Chattopadhyay:2021eba}
D.~S.~Chattopadhyay, K.~Chakraborty, A.~Dighe, S.~Goswami and S.~M.~Lakshmi,
Phys. Rev. Lett. \textbf{129} (2022) no.1, 011802
[arXiv:2111.13128 [hep-ph]].










\bibitem{Chikashige:1980ui}
Y.~Chikashige, R.~N.~Mohapatra and R.~D.~Peccei,
Phys. Lett. B \textbf{98}, 265-268 (1981).

\bibitem{Gelmini:1980re}
G.~B.~Gelmini and M.~Roncadelli,
Phys. Lett. B \textbf{99}, 411-415 (1981).

\bibitem{Kim:1990km}
C.~W.~Kim and W.~P.~Lam,
Mod. Phys. Lett. A \textbf{5}, 297-299 (1990).

\bibitem{Ando:2004qe}
S.~Ando,
Phys. Rev. D \textbf{70}, 033004 (2004),
[arXiv:hep-ph/0405200 [hep-ph]].

\bibitem{Strumia:2003zx}
A.~Strumia and F.~Vissani,
Phys. Lett. B \textbf{564} (2003), 42-54,
[arXiv:astro-ph/0302055 [astro-ph]].

\bibitem{Ricciardi:2022pru}
G.~Ricciardi, N.~Vignaroli and F.~Vissani,
JHEP \textbf{08}, 212 (2022),
[arXiv:2206.05567 [hep-ph]].

\bibitem{Vissani:2014doa}
F.~Vissani,
J. Phys. G \textbf{42}, 013001 (2015),
[arXiv:1409.4710 [astro-ph.HE]].

\bibitem{Cowan:2010js}
G.~Cowan, K.~Cranmer, E.~Gross and O.~Vitells,
Eur. Phys. J. C \textbf{71}, 1554 (2011),
[erratum: Eur. Phys. J. C \textbf{73}, 2501 (2013)],
[arXiv:1007.1727 [physics.data-an]].



\bibitem{Lunardini:2004bj}
C.~Lunardini and A.~Y.~Smirnov,
Astropart. Phys. \textbf{21}, 703-720 (2004)
[arXiv:hep-ph/0402128 [hep-ph]].

\bibitem{Jegerlehner:1996kx}
B.~Jegerlehner, F.~Neubig and G.~Raffelt,
Phys. Rev. D \textbf{54}, 1194-1203 (1996),
[arXiv:astro-ph/9601111 [astro-ph]].

\bibitem{Porto-Silva:2020gma}
Y.~P.~Porto-Silva, S.~Prakash, O.~L.~G.~Peres, H.~Nunokawa and H.~Minakata,
Eur. Phys. J. C \textbf{80} (2020) no.10, 999
[arXiv:2002.12134 [hep-ph]].

\bibitem{Funcke:2019grs}
L.~Funcke, G.~Raffelt and E.~Vitagliano,
Phys. Rev. D \textbf{101} (2020) no.1, 015025
[arXiv:1905.01264 [hep-ph]].

\bibitem{Picoreti:2021yct}
R.~Picoreti, D.~Pramanik, P.~C.~de Holanda and O.~L.~G.~Peres,
Phys. Rev. D \textbf{106} (2022) no.1, 015025
[arXiv:2109.13272 [hep-ph]].




\end{thebibliography}
\end{document}